 \documentclass[aps,a4paper,twocolumn,showpacs]{revtex4}

\usepackage{graphicx}

\begin{document}

\newcommand{\be} {\begin{equation}}
\newcommand{\ee} {\end{equation}}
\newcommand{\ba} {\begin{eqnarray}}
\newcommand{\ea} {\end{eqnarray}}
\newcommand{\tr} {{\rm tr}}

\title{Semiclassical Propagation of Gaussian Wavepackets} 

\author{Raphael N. P. Maia}

\author{Fernando Nicacio}

\author{Ra\'ul O. Vallejos}
%
\affiliation{ Centro Brasileiro de Pesquisas F\'{\i}sicas (CBPF), 
              Rua Dr.~Xavier Sigaud 150, 
              22290-180 Rio de Janeiro, Brazil}

\author{Fabricio Toscano}
%
\affiliation{ Funda\c{c}\~ao Centro de Ci\^encias e 
              Educa\c{c}\~ao Superior a Dist\^ancia do Estado 
              do Rio de Janeiro,
              20943-001 Rio de Janeiro, Brazil 
              \\ and Instituto de F\'{\i}sica, 
              Universidade Federal do Rio de Janeiro, 
              Cx.~P. 68528, 21941-972 Rio de Janeiro, Brazil}

\date{\today}

\begin{abstract}
We analyze the semiclassical evolution of Gaussian wavepackets 
in chaotic systems.
We 
prove that after some short time a Gaussian wavepacket 
becomes a primitive WKB state.
From then on, the state can be propagated using the standard
TDWKB scheme.
Complex trajectories are not necessary to account for the 
long-time propagation.
The Wigner function of the evolving state develops the structure 
of a classical filament plus quantum oscillations, with phase and 
amplitude being determined by geometric properties of a
classical manifold.
\end{abstract}



\maketitle

%

{\em Introduction}.
%
The standard approach to semiclassical evolution is
time-dependent WKB theory (TDWKB) 
\cite{vanvleck28,dirac,berry79,maslov}. 
This theory provides a clear geometric description of the
dynamics: a time dependent quantum state is associated with 
an evolving Lagrangian manifold in classical phase space. 
[A phase-space manifold $(p(q),q)$ is Lagrangian if 
$p=\nabla S(q)$, for some generating function $S$ 
\cite{littlejohn92}.]
For TDWKB to be applicable the initial state 
must itself be related to a Lagrangian manifold.
This is the case, e.g., of eigenstates of position or momentum 
operators (related to planes) \cite{littlejohn92},
or highly excited eigenstates of bounded integrable Hamiltonians 
(related to tori) \cite{berry79}.

Somewhat surprisingly TDWKB has never been used
to propagate Gaussian wavepackets.
It is probably the static point of view what blocked the 
use of standard TDWKB:
If a Gaussian state is thought of as the ground state of
some harmonic oscillator, then it certainly does not qualify 
as an initial WKB state.
However, by taking the dynamics into consideration,
a new perspective arises.
In chaotic systems, when one observes the evolution
of an initially Gaussian wavepacket through a phase space 
representation (like Husimi or Wigner) \cite{schleich}, 
it becomes manifest that, after some time, the wavepacket 
acquires the form of a thin filament, very similar to
the classical evolution of the initial density 
\cite{tomsovic91,zurek03,silvestrov03a,toscano05}
(in the case of the Wigner function the filament is decorated
by interference fringes).
The smaller $\hbar$ (as compared with the relevant action 
scales), the stronger the localization of the wavepacket
along some classical manifold. 
With this picture in mind it is natural to conjecture that 
the wavepacket evolves into a WKB state, its 
support being a real phase-space manifold 
\cite{silvestrov02,silvestrov03a,schubert04}.
The purpose of this paper is to prove this statement.

We show 
that, after some (short) time, a Gaussian wavepacket 
becomes a primitive WKB state.
From then on, the state can be propagated using the standard
TDWKB scheme.
Complex trajectories \cite{maslovII,huber88,aguiar05} are not 
necessary to describe the long-time propagation of wavepackets,
but they may be used to describe the evolution during
the initial stage.
The present approach not only offers an intuitive geometric 
description of the evolving state, but can be very accurate,
as we demonstrate with a numerical example.

We focus on the Wigner function for its
ability to reflect subjacent phase space structures 
\cite{ozorio98}, and because it is the standard representation 
when dealing with decoherence (arising from coupling to the 
environment \cite{zurek03}) in semiclassical regimes.

\vspace{1pc}
{\em TDWKB approach}.
%
In order to eliminate some unnecessary complications in 
the general problem of semiclassical wavepacket propagation 
in chaotic systems we consider a one degree of freedom 
Hamiltonian $H(p,q,t)$, the dependence on $t$ being periodic 
with period $\tau$. 
Let us assume that the one-period map,
$ M_\tau(p(t),q(t))=(p(t+\tau),q(t+\tau)) $,
has a hyperbolic fixed point at the origin, and, without 
loss of generality, choose the $q$ axis to coincide with 
the unstable subspace.
At $t=0$ we launch a Gaussian wavepacket at the origin:
\be
\label{initial}
\psi_0(q)= (2 \pi \sigma^2)^{-1/4}  
           \exp (-q^2 / 4 \sigma^2) \; .
\ee
Note that we have preserved the essential ingredient of
chaotic dynamics, i.e., 
the exponential stretching and folding of phase space 
manifolds. 

If the $q$-uncertainty ($\sigma$) of the initial 
wavepacket (\ref{initial}) is large enough 
[e.g., a classical length, $\sigma \sim \hbar^0$],
then (\ref{initial}) is already a primitive WKB state, 
\be
\label{initialWKB}
\psi_0(q)= A_0(q)  \exp [ i S_0(q) / \hbar ] \; ,
\ee
given that both the amplitude $A_0(q)$ and phase 
$S_0(q)=0$ vary slowly on the quantum scale \cite{dirac}. 
The associated Lagrangian manifold is $p=0=dS_0/dq$.
Accordingly, this state can be successfully evolved 
using the TDWKB scheme, as we showed in Ref.~\cite{maia07}.

\begin{figure}[htp]
\hspace{0.0cm}
\includegraphics[angle=-90, width=8cm]{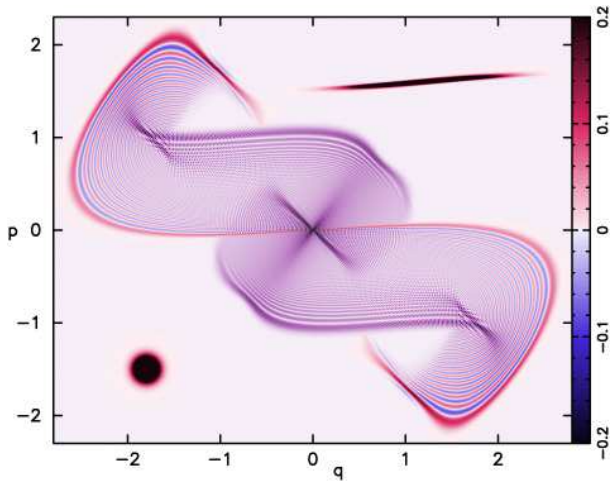}
\caption{%
(Color online) 
Linear density plot of the exact Wigner function 
[$\pi \hbar \, W(p,q)$] after six periods of evolution 
with the kicked harmonic oscillator 
(see the text for a description of the system). 
We also show displaced versions of 
(i) the initial state 
(a circular Gaussian at the origin, left-bottom), 
(ii) the Wigner function after two steps (top-right).}
\label{fig1}
\end{figure}

Let us now analyze what happens when a small circular 
wavepacket is launched at an unstable fixed point.
Numerical simulations show that the positive part of the 
Wigner function gets stretched along the unstable 
manifold. 
As this positive part bends, interference fringes appear. 
The picture is that of a positive (classical) thin filament 
decorated by an oscillatory pattern (see Fig.~\ref{fig1}).
%
%
%
%

The fact that the structures in Fig.~\ref{fig1} are very 
similar to those found in the Wigner functions related 
to WKB eigenstates \cite{berry77}, leads us to enquire: 
does the Wigner function of Fig.~\ref{fig1} correspond to 
a WKB state, i.e., can we write
\be
\label{WKB}
\psi_t(q) 
\approx \sum_\nu              A^{(\nu)}_t(q)  
                  \exp [ \, i S^{(\nu)}_t(q) / \hbar ]  \;?
\ee
Here $\nu$ labels the different branches of the 
hypothetical Lagrangian manifold supporting the WKB state 
($p^{(\nu)}=dS^{(\nu)}/dq$) \cite{littlejohn92}. 
It is understood that the expression above must be valid
during some time interval, during which amplitudes and 
phases evolve according to TDWKB theory
\cite{dirac,berry79,maslov,littlejohn92}, i.e.,
amplitudes are convected by the classical flow in
$q$-space 
\be
A_t(q') = A_0(q) 
\left| \partial q'/\partial q \right|^{-1/2} \;,
\ee
and the generating function, which solves the 
Hamilton-Jacobi equation, can be written as
an integral over the classically evolved manifold 
\cite{berry79}:
\be
S_t(q') = S_t (q_0) + \int_{q_0}^{q'} p \, dq  \;.
\ee

In the case of only one branch, i.e., before the 
classical manifold folds, a numerical simulation
is sufficient to answer the question posed above, 
as we show in the following.
For this purpose we introduce the concrete model
system that will serve as our test bench.
This is the kicked harmonic oscillator (KHO) 
\cite{berman91,zaslavsky,toscano05}:
\be
H(p,q,t)= \frac{p^2}{2m} + \frac{1}{2} m \omega^2 q^2 +
  K \cos (kq) \sum_{n=0}^\infty \delta( t - n\tau) \,.
\ee
The parameters $m=\omega=k=1$, $K=2$ 
(in appropriate units \cite{toscano05}), 
and $\tau=\pi/(3\omega)$ guarantee a large chaotic region 
around the hyperbolic fixed point located at the origin 
\cite{toscano05}.
The unstable direction is close to the $q$ axis.
We considered an initial circular state ($t=0^-$) given by 
Eq.~(\ref{initial}) with $\sigma= 0.08$ 
(corresponding to $\hbar=0.0128$). 
Some snapshots of its evolution (Wigner functions) are shown 
in Fig.~\ref{fig1}.
%
%
\begin{figure}[htp]
\hspace{0.0cm}
\includegraphics[angle=0, width=9cm]{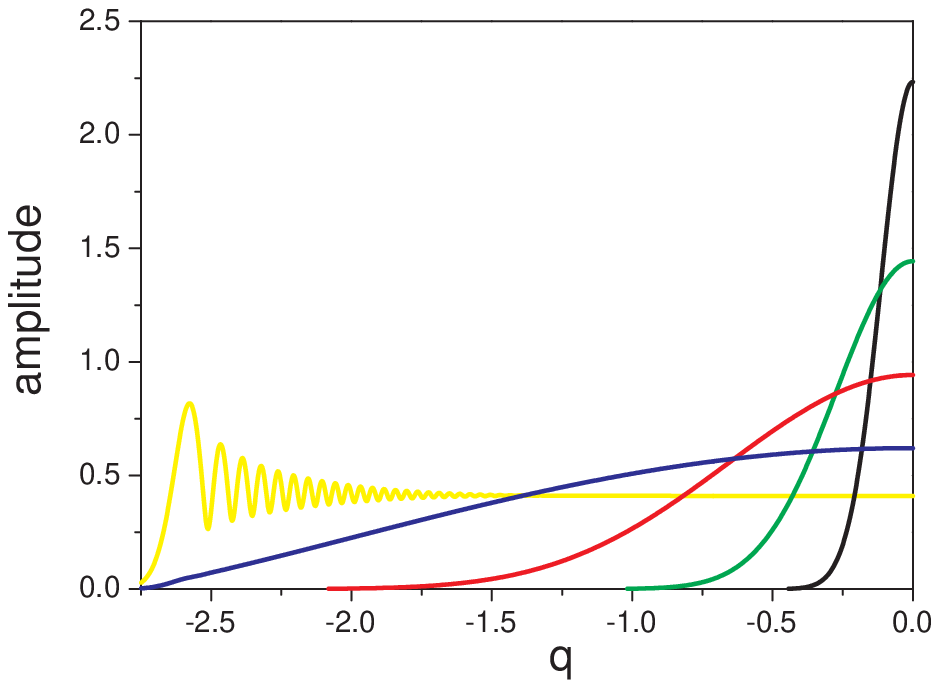}
\vskip-6.3cm
\includegraphics[angle=0, width=5cm]{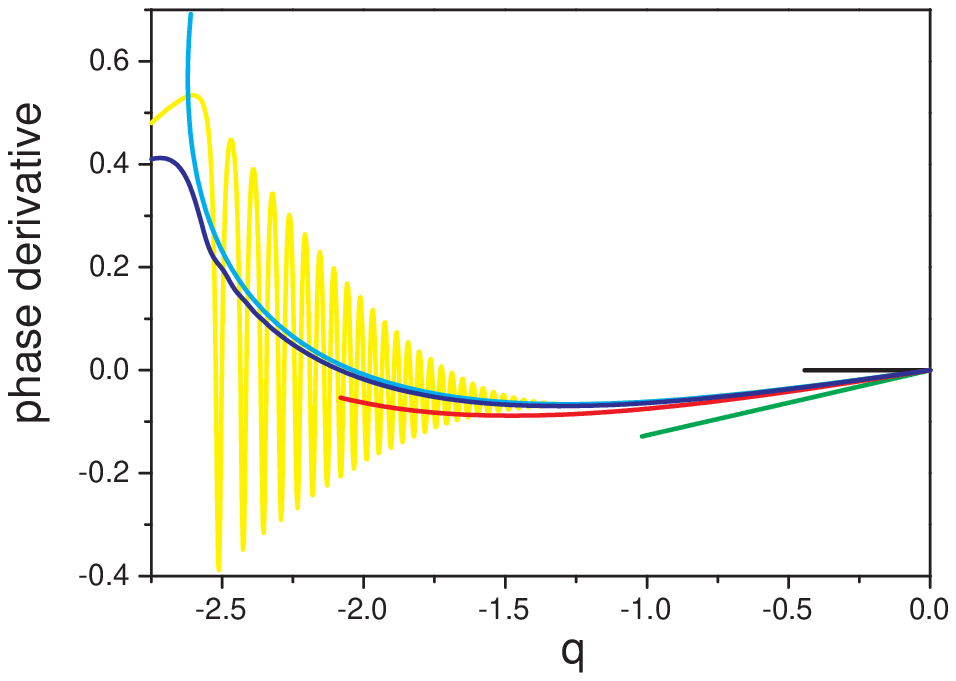}
\vskip+2.0cm
\caption{%
(Color online) 
Wavefunction amplitude $|\psi_t(q)|$ and
phase derivative $d\phi_t/dq$ (inset) at $t/\tau=0,1,2,3,4$
(black, green, red, blue, yellow, respectively).
Also shown is the unstable manifold (inset, cyan). 
Due to parity symmetry we only plot the region 
$q \le 0$.}
\label{fig2}
\end{figure}
%
%

In Fig.~\ref{fig2} we show amplitude $\rho_t(q)$ 
and phase derivative $d\phi_t/dq$ of the evolved 
state for short times
[$\psi \equiv \rho \exp(i\phi/\hbar)$]. 
Both quantities must be smooth on the quantum scale 
for the state to qualify as a primitive WKB state. 
We see that as time grows the amplitude gets smoother: 
at $t=0$ it was localized in a region of size 
${\cal O}(\hbar^{1/2})$, 
at $t=3$ it has acquired the maximal (classical) width.
The derivative of the phase, which is the candidate to 
Lagrangian support of the WKB state,
stays smooth, and close to the unstable manifold, 
until $t \approx 3$. 
For larger times, $t \ge 4$, the turning point is 
reached and a fold is born, giving rise to 
quantum oscillations in both $\rho(q)$ and $\phi(q)$. 
For smaller $\hbar$ a similar behavior is observed,
the only difference being that it takes longer to reach
the turning point (this time goes like $\log \hbar$).
Our claim is that, for small enough $\hbar$, there 
is a time window 
$(t_{\rm min},t_{\rm max})$
where $\psi_t$ is to good accuracy (see below) a 
primitive WKB state, meaning that it can be 
propagated further on according to TDWKB.
In our numerical example, we checked that the 
optimal time for starting the TDWKB scheme is $t=2$.
At $t=3$ the interference effects of the turning point are 
already significant (even if not apparent in 
Fig.~\ref{fig2}). 
At $t=1$ the wavepacket is still not wide enough. 

Incidentally, the inset of Fig.~\ref{fig2} shows
that TDWKB performs badly if applied at $t=0$.
The initial manifold $p=0$ does not evolve 
classically:
it jumps over the unstable direction,
instead of staying on the same side.

The next step is thus to evolve $\psi_{t=2}$ 
using TDWKB and compare with the exact propagation.
In order to unfold subjacent phase-space structures,
the comparison will be made at the level of Wigner 
functions.

\begin{figure}[htp]
\hspace{0.0cm}
\includegraphics[angle=0, width=9cm]{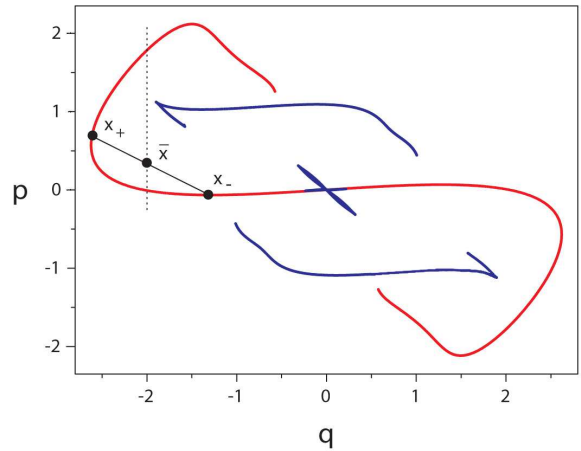}
\caption{(Color online) Classical manifold corresponding
to the state depicted in Fig.~\ref{fig1} (red). 
This manifold is itself a caustic of the Wigner function.
The ``ghost" lines (blue) also belong to the caustic.
The point $\bar{x}$ is the center of the chord with tips at 
$x_-$ and $x_+$.
A point $\bar{x}$ belongs to the caustic if the vectors tangent 
to the manifold at $x_-$ and $x_+$ are parallel.
The dashed vertical line at $q=-2.0$ indicates a special 
section for testing the semiclassical approximations
(see Fig.~\ref{fig4}).}
\label{fig3}
\end{figure}

\vspace{1pc}
{\em Semiclassical Wigner function}.
%
The Wigner function of a general WKB state 
[Eq.~(\ref{WKB})] can be calculated analytically, 
using the stationary phase method,
in a way analogous to that followed by Berry and Balazs 
for the special case of semiclassical eigenstates 
of integrable Hamiltonians \cite{berry79}.
We start with the usual definition of Wigner function, 
specialized to our case,
\be
\label{integral}
W(\bar{p},\bar{q},t)=
\frac{1}{\pi \hbar} \int d\xi A_t(\bar{q}+\xi) A_t(\bar{q}-\xi)
e^{ i \phi(\xi) / \hbar} \; ,
\ee
where the phase is given by
$ 
\phi(\xi) = S_t(\bar{q}+\xi) - S_t(\bar{q}-\xi) - 2 \bar{p} \xi
$.
We assume that $(\bar{p},\bar{q})$ is not far away from the manifold.
In this case only one branch of $S_t$ is relevant \cite{berry79}.
The stationary phase condition reads
$
p_t(\bar{q}+\xi) + p_t(\bar{q}-\xi) = 2 \bar{p}
$,
where $p_t(q)=dS_t/dq$.
If the point $\bar{x}=(\bar{p},\bar{q})$ is on the concave side of 
the evolved manifold but not too far away, there are in general
two solutions $\pm \xi^\ast$, defining two points on the
manifold, $x_+$ and $x_-$. 
These are the tips of a chord having $\bar{x}$ as midpoint 
(Fig.~\ref{fig3}).
When the stationary points are not coalescing they give
individual complex conjugate contributions to the integral.
The corresponding Wigner function reads
\be
\label{wigner}
W(\bar{p},\bar{q},t)=
\frac{2\sqrt{2}} {\sqrt{\pi \hbar}} A_0(q_+) A_0(q_-)  
\frac{\cos\left({\cal A}/\hbar - \pi/4 \right)}
     {\sqrt{|v_+ \wedge v_-|}} \; .
\ee
The phase structure was extracted literally from 
Ref.~\cite{berry79}, i.e., 
${\cal A}$ is the (symplectic) area between the manifold
and the chord.
The amplitude is different from Berry and Balazs',
as they considered a different initial density. 
Here $A_0(q_\pm)$ is the initial 
amplitude at the preimages of $x_\pm$ and
$
v_\pm = d x_\pm / dq_0 \,
$
denote tangent vectors at $x_\pm$, their moduli representing 
local rates of expansion ($dq_0$ is the $q$ component of 
a displacement along the initial manifold, at the preimage 
of $x_\pm$).

Equation~(\ref{wigner}) is a good semiclassical approximation,
except in the vicinity caustic points, where 
$|v_+ \wedge v_-|=0$, 
i.e., when tangent vectors at the tips of the chords are 
parallel (see Fig.~\ref{fig3}).
%
%
At caustics stationary phase points coalesce and one must 
use transitional (or uniform) approximations \cite{berry77}. 
If $(\bar{p},\bar{q})$ is close enough to the classical
manifold, we can obtain a crude transitional approximation
to Eq.~(\ref{integral}) as follows: 
(i) Approximate the manifold by the quadratic curve
$\bar{p} \approx 
p_\ast + p'_\ast(q-q_\ast)+p^{''}_\ast(q-q_\ast)^2/2$.
This leads to a cubic phase $\phi(\xi)$.
(ii) Neglect variations of the amplitude, 
i.e., $A_t(q)\approx A_t(q_\ast)$. 
If we traverse the caustic along
the line $\bar{q}=q_\ast$, and assuming $p^{''}_\ast>0$, then 
\cite{alonso00,silvestrov02}:
\be
\label{trans}
W(\bar{p},q_\ast,t) 
\approx
\frac{2 A_t^2(q_\ast)}{ (\hbar^2 p^{''}_\ast)^{1/3}} 
{\rm Ai} \left[ -\frac{2(\bar{p}-p_\ast)}
                     {(\hbar^2 p^{''}_\ast)^{1/3}} \right] \, ,
\ee
Ai standing for the Airy function.

We are now in position for the final step in our program:
the comparison of exact and semiclassical evolutions.
Figure~\ref{fig1} displays the exact Wigner function after
six steps of evolution.
It is evident that its skeleton is the caustic of the 
manifold of Fig.~\ref{fig3}, which was obtained by 
evolving during four steps the manifold (phase derivative) 
associated to the exact $\psi_{t=2}(q)$.

\begin{figure}[htp]
\begin{center}
\includegraphics[angle=0, width=9cm]{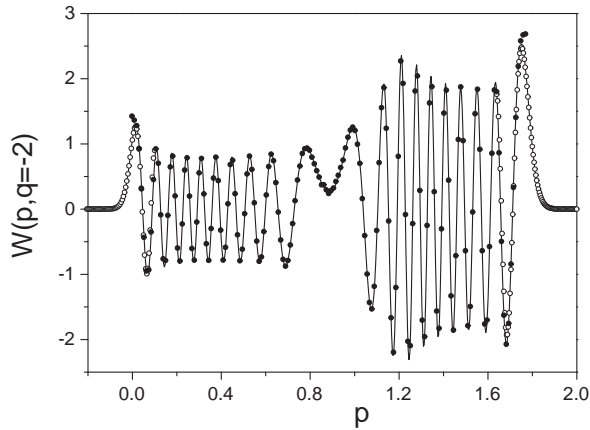}
\caption{%
A section of the exact Wigner function $W(p,q=-2)$ 
of Fig.~\ref{fig1} (line) vs. WKB approximation. 
Full circles correspond to the standard stationary
phase result [Eq.~(\ref{wigner})] and open circles to 
the transitional approximation [Eq.~(\ref{trans})]. }
\label{fig4}
\end{center}
\end{figure}

For a quantitative examination, in Fig.~\ref{fig4}
we plotted the section $q=-2$ of the exact Wigner 
function together with the semiclassical prediction, 
Eqs.~(\ref{wigner}) and (\ref{trans}).
Inside their respective domains of validity both 
approximations are excellent.
This completes our argumentation.

\vspace{1pc}
{\em Concluding remarks}.
%
We reported for the first time the use of standard TDWKB
for semiclassical propagation of wavepackets in chaotic 
systems.
 
The key point is that chaotic dynamics provides 
the initial expansion that defines the appropriate Lagrangian 
manifold for starting the TDWKB scheme.
In particular we showed that localized states typically
evolve into WKB states, and explained how to calculate
amplitudes and phases explicitly.

%
%

Our results provide a novel perspective for re-examining 
important previous work on long-time wavepacket propagation.
Consider, for instance, the remarkable calculations
of Tomsovic and Heller's, who used a multiple 
linearization scheme to obtain accurate 
autocorrelation functions for large times \cite{tomsovic91}.
In the light of our findings, their scheme can now be 
understood as arising from the linearization of the WKB 
wavefunction $\psi_t(q)$ (which is globally valid) 
in the vicinity of a periodic point.
The family of ``homoclinic" intersections in 
Ref.~\cite{tomsovic91}, essential for organizing the 
summation of recurrences, corresponds, in the TWKB context,
to the set of intersections between the 
Lagrangian manifold of the evolved state and 
the stable manifold of the fixed point.
In this way, we expect the present paper will contribute to 
the ongoing debate about the timescale for the breakdown of 
semiclassical propagation.
Whether the breaktime diverges like $\log \hbar$ 
\cite{silvestrov02,schubert04} or, more plausibly, like some 
power of $\hbar$ \cite{tomsovic91,schubert07} is a question 
still awaiting a definitive answer.

One important feature of TDWKB is that it can be easily 
supplemented to accommodate decoherence effects.
Let us briefly analyse the example of the Lindblad master 
equation corresponding to a chaotic Hamiltonian system
coupled to a high temperature reservoir 
\cite{breuer,toscano05}
(or, similarly, the non-selective weak continuous measurement 
of $\hat q$ and/or $\hat p$ \cite{breuer,jacobs06}).
In the formalism of quantum trajectories \cite{breuer}, 
the quantum jumps associated to such an environment 
(or weak measurement scheme), amount to random rigid 
translations in phase space \cite{hall94}. 
The alternation of random translations with 
Hamiltonian evolution leads to a final density matrix 
represented by a weighted ensemble of pure WKB states, each one 
related to a particular history of random translations.  
The corresponding Wigner function is thus suggestively expressed
as an average over filamentary Wigner functions like the one 
depicted in Fig.~\ref{fig1}.

\vspace{1pc}
{\em Acknowledgments}.
%
We thank 
    A. M. Ozorio de Almeida, 
    L. A. Pach\'on,
    M.    Sieber,
    E.    Vergini,
    T.    Dittrich,
    J.    Vanicek,
    L.    Kaplan, 
    G.    Alber,
and 
    E. J. Heller
for interesting comments.
Partial financial support from 
    CNPq, 
    CAPES, 
    PROSUL, 
    The Millennium Institute for Quantum Information 
    (Brazilian agencies), 
and 
    UNESCO/IBSP Project 3-BR-06 
is gratefully acknowledged. 
%


\end{document}